\documentclass[11pt,a4paper,copyright,goog]{google}

\usepackage[numbers,compress]{natbib}
\usepackage{xurl}
\usepackage{hyperref}
\usepackage{cleveref}
\usepackage{multirow}
\usepackage{bm}
\usepackage{quoting}
\usepackage{ragged2e}
\usepackage{array}
\usepackage{enumitem}
\usepackage{rotating}

\crefformat{footnote}{#2\footnotemark[#1]#3}

\newcommand{\quoteinline}[1]{\sffamily{``#1''}}
\newcolumntype{C}[1]{>{\centering\let\newline\\\arraybackslash\hspace{0pt}}m{#1}}

\usepackage{booktabs}      
\usepackage{array}         
\usepackage{caption}       
\usepackage{graphicx}
\usepackage{subcaption}
\usepackage{chngcntr}
\usepackage{longtable}
\usepackage{listings}
\usepackage{xcolor}

\definecolor{backcolour}{rgb}{0.95,0.95,0.92}

\lstdefinestyle{monospacedgrey}{
    backgroundcolor=\color{backcolour} \& 
    basicstyle=\ttfamily\scriptsize \& 
    numbers=none \& 
    breaklines=true \& 
    breakatwhitespace=true \& 
    postbreak=\mbox{\textcolor{red}{$\hookrightarrow$}\space} \& 
}

\newcommand{\takeout}[1]{}

\newcommand{\xPlain}{Learn Your Way}
\newcommand{\myHeader}[1]{\noindent {\bf #1}.\\}
\newcommand{\figref}[1]{Figure~\ref{#1}}

\newcommand{\appref}[1]{Appendix~\ref{#1}}

\newcommand{\ag}[1]{{\color{red}{AG: {#1}}}}

\newcommand{\yh}[1]{}

\newcommand{\agbib}[1]{}

\newcommand{\secref}[1]{Section \ref{#1}}
\newcommand{\commentout}[1]{}

\paperurl{}

\uselogo{} 

\title{Towards an AI-Augmented Textbook}
\author[]{LearnLM Team,  Google}

\begin{abstract}
Textbooks are a cornerstone of education, but they have a fundamental limitation: they are a one-size-fits-all medium. Any new material or alternative representation  requires arduous human effort, so that textbooks cannot be adapted in a scalable manner. We present an approach for transforming and augmenting textbooks using generative AI, adding layers of multiple representations and personalization while maintaining content integrity and quality. We refer to the system built with this approach as \xPlain{}. We report pedagogical evaluations of the different transformations and augmentations, and present the results of a a randomized control trial, highlighting the advantages of learning with \xPlain{} over regular textbook usage.
\end{abstract}

\begin{document}
\maketitle

\section{Introduction}

Recent advances in generative Artificial Intelligence (Gen-AI) have the potential to revolutionize education, but this potential is yet to be realized in full. It requires a responsible, multidisciplinary approach to weave together learning science and cutting edge technology. In this work, we focus on a central aspect of the current learning journey: exploring textbook material. Traditionally, every school selects several textbooks that are meant for use by all learners. The textbooks, by definition, are inflexible and not adaptive, as it is impractical to manually create a version for every audience, and certainly not one that would adapt to individual user needs. 
Here we argue that in the age of Gen-AI, this
notion of a flexible and personalized textbook is in fact within reach. Specifically, we show how textbooks can be transformed into a richer and more personalized form, while maintaining the integrity of the original content, and adding  layers that promote effective learning.


\begin{figure}[t]
\centering
\vspace{-0.15in}
\includegraphics[trim={0 10 0 0},clip,width=0.9\columnwidth]{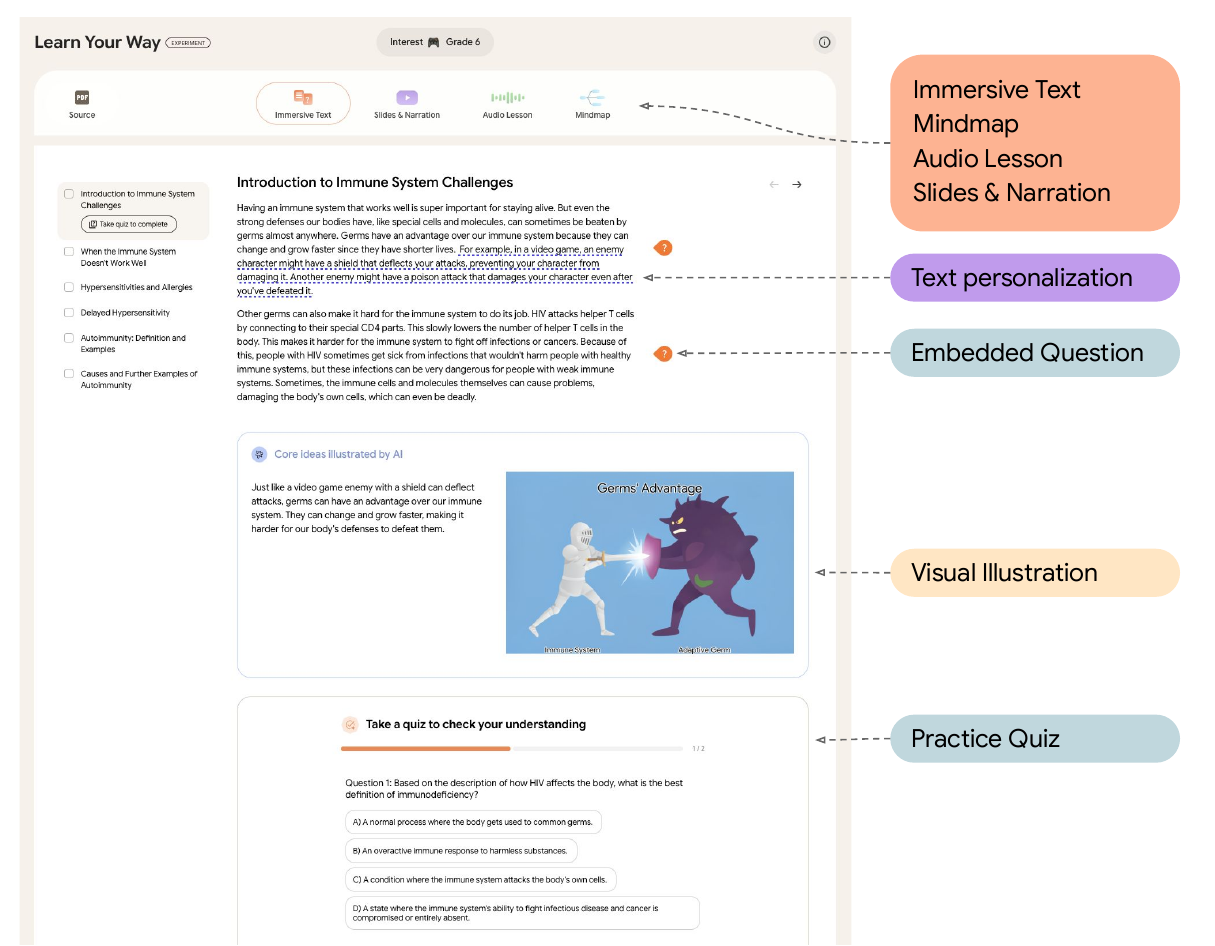}
\caption{An example of the \xPlain{} learning experience. Centerpiece is the ``Immersive Text'' view, that shows the 
source material (OpenStax's \emph{Disruptions in the Immune System} content) transformed to 6th grade level and adapted for a personal interest in gaming. The Immersive Text contains various generative add-ons such as personalized examples, embedded questions, and more. At any given time, the learner can also switch to alternative views of the entire material such as narrated slides or an audio lesson, which are also personalized.}
\label{fig:main}
\end{figure}

Our textbook augmentation approach takes as input a textbook segment or chapters  and uses them as the basis for extensive generated content, practice and evaluation. Our approach rests on two key concepts that underlie the corresponding augmentations of the original content: multiple representations and personalization. 
We propose a two step AI generation scheme whereby the original text is first personalized, and then transformed into a range of presentation forms and assessment components. A key desiderata in this process is that content is adequately aligned with the source and curriculum, and that the presentation is engaging and pedagogically effective. We implement our approach in an experimental learning experience that we call \xPlain{}. 

We begin with the pedagogical observation that learning can be more effective when the experience is adapted to the characteristics and needs of the learner \citep[see][for a review of personalization approaches]{bernacki2021systematic,shemshack2020systematic}.
\xPlain{} is thus designed to first re-generate the original textbook content, based on specific learner attributes. In addition, \xPlain{} generates assessment opportunities for the learner which serve to create a signal about their progress, reflect personalized feedback to the learner, and influence subsequent steps.

The value of multiple representations of content has been studied in learning science \citep[e.g., see][]{AINSWORTH1999131}. For example, dual coding theory \cite{clark1991dual} suggests that multiple representations have the advantage of forging links between different encodings of the same concepts, thus reinforcing the corresponding mental conceptual structures. 
\xPlain{} is therefore augmented with multiple views of the material (audio lessons, narrated slides, and mind maps), which learners can interact with and choose from.
Providing these options is in line with the view that personalized systems should also be adaptable, offering learners agency to decide on their learning path \cite{CHERNIKOVA2025100662,aleven2016instruction,plass2020toward}. 
It is also motivated by theories of self-regulated learning (SRL) \cite{zimmerman2011self} and visible learning \cite{hattie2008visible} that acknowledge the importance of supporting learners with cognate control of their learning process.

\figref{fig:main} shows the central \xPlain{} view, demonstrating how personalization and multiple representations come together. The resulting AI-augmented textbook provides the learner with a personalized and engaging learning experience, while also allowing them to choose from different modalities in order to enhance understanding. 
In the next sections we describe each of the components in \xPlain{}, along with an evaluation of the pedagogical merit of each one. Finally, we report the results of a randomized controlled study showing that learning with our  personalization and multiple representations \xPlain{} system can improve learning efficacy compared to a standard Digital Reader over the same material. Taken together, our results demonstrate the potential to re-imagine the medium of a textbook in the age of generative AI. 

\section{Textbook Augmentation via Personalization and Multiple-Views}

We assume source-of-truth material which is defined by the learner’s curriculum and learning goals. For simplicity, think of a section in a textbook, although the source-of-truth can be a more complex collection of knowledge and skills to be delivered. Our goal is to explore how transforming the source material can increase content engagement and efficacy. Gen-AI offers four key opportunities in this context. First, it can generate such content for any material the learner is interested in. Second, it can do so while adapting to the specific attributes and needs of the learner. This is in contrast to the generation of personalized learning material by human educators, which is a much longer process and is impractical to do at scale. Third, AI can be used to generate different representations of the material, including visualizations and audio-based formats, which are known to further enhance the efficacy of learning \cite{AINSWORTH1999131,clark1991dual}. Finally, AI can generate formative assessment elements tailored to the learner, allowing them to monitor and regulate progress. As \cite{clark2012formative} notes, formative assessment is a critical driver of learning, and in particular self-regulated learning.




\begin{figure}[ht!]
    \centering 
    \includegraphics[trim={0 0 0 1.2in},clip,width=0.95\linewidth]{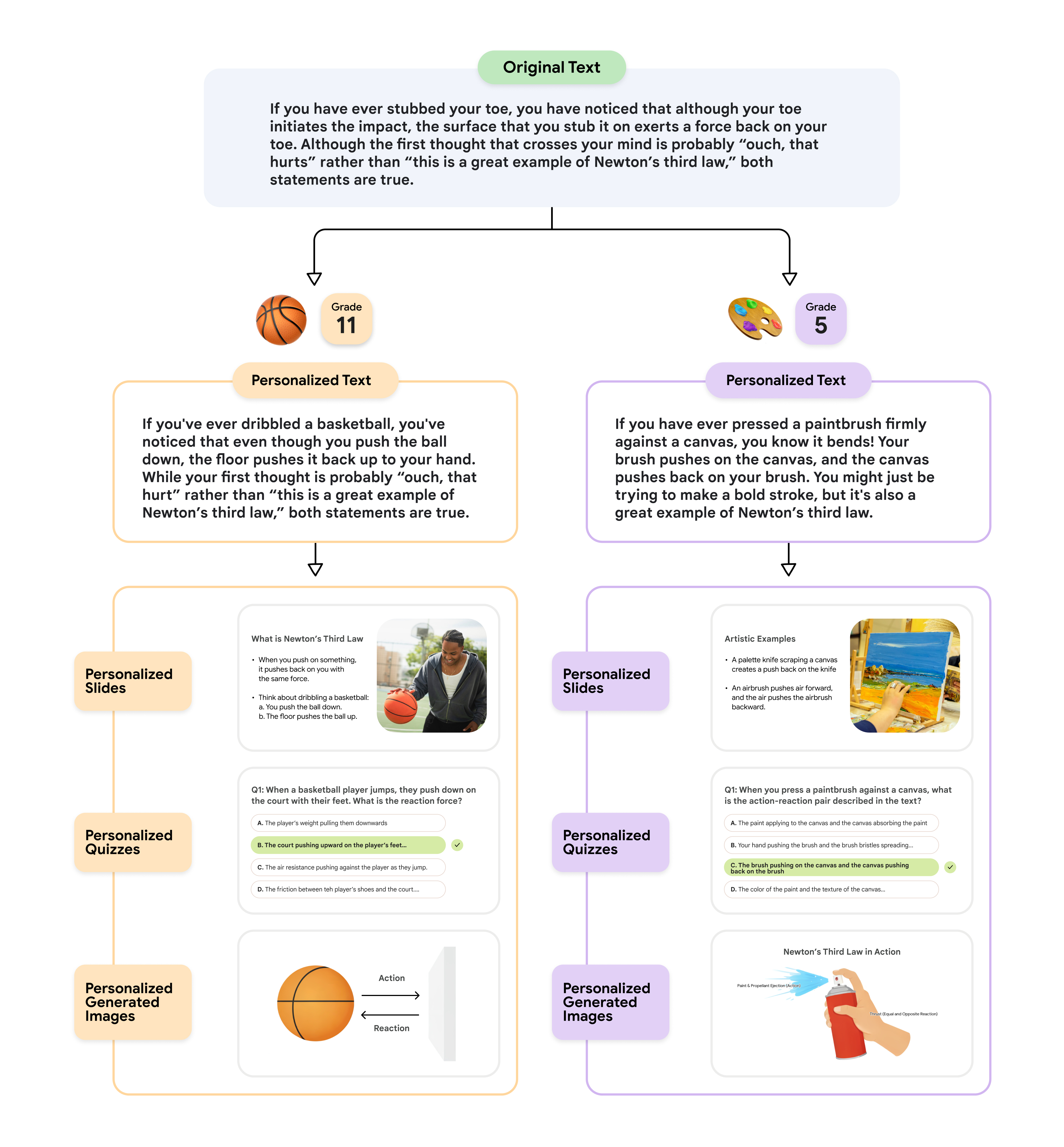}
    \caption{An illustration of the two step generation procedure used in \xPlain{}. Here, a generic example from OpenStax's \emph{Newton's Third Law of Motion} is first personalized for the personal interest of 'basketball' (left) and 'art' (right), and then expanded into different presentation formats.}
    \label{fig:personalization_then_expansion}
\end{figure}

Our textbook transformation and augmentation follows a two step approach. In the ``Text Personalization'' stage, we rewrite the material to match specific personal attributes of the learner. Then in the ``Content Transformations'' stage, we create multiple views of the rewritten material. These allow the user to choose their own learning path, interleaving complementary representations of the same conceptual structures.
\figref{fig:personalization_then_expansion} demonstrates this process, showing two different personalization transformations, that results in different views.
Unless otherwise noted, all transformations and augmentations described below rely directly on Gemini 2.5 Pro, a leading model for education \cite{Arena}, without additional fine-tuning.



\subsection{Text Personalization \label{sec:text_rewrite}}
As explained above, our approach first transforms the original text into a more personalized form. A key choice in this process is what specific attributes of the learner should be personalized to. For simplicity, we focus on two key attributes: grade-level and personal interests. There are of course many additional attributes to consider on the path towards more comprehensive personalization.

\myHeader{Personalization to Grade Level}
Adaptation of the material to match the reading grade level of the learner is a core transformation that provides the basis for all other transformations that follow. The text is generatively adapted, with the goal of matching the Flesch-Kincaid Grade (FKG) \cite{Kincaid1975DerivationON,kincaid1975derivation} for that level, while maintaining factuality and coverage of the material. This is known as re-leveling and is part of Gemini 2.5 Pro core education capabilities. See \cite{Arena}, Section 2.2 for an evaluation.

\myHeader{Personalization to Interests}
The \xPlain{} experience asks the learners, in addition to the grade level, for their personal interests. Currently, for simplicity, the learner is asked to select one of several common interests (e.g., sports, music, food).
This information is then used to rewrite the original text, making it more relatable. This also serves the purpose of mapping new knowledge to existing conceptual networks used by the learners, thus making learning more effective. As \cite{hattan2024leveraging} notes: {\quoteinline{individuals’ existing knowledge serves as a base for subsequent learning and performance}}. Their review further argues that {\quoteinline{prior knowledge guides readers’ comprehension of written language}}. Our Gen-AI rewriting is done in a focused manner, by first selecting parts of the text that are particularly amenable to personalization, and then replacing only these parts with an AI-rewritten personalized version. This has the added advantage of highlighting the personalized text, thus informing the learner that it has been specialized to their interests. See example in \figref{fig:personalization_then_expansion} for Newton's third law example, rewritten for two different interests: basketball and art.


\begin{figure}[t]
\centering
\includegraphics[width=0.75\columnwidth]{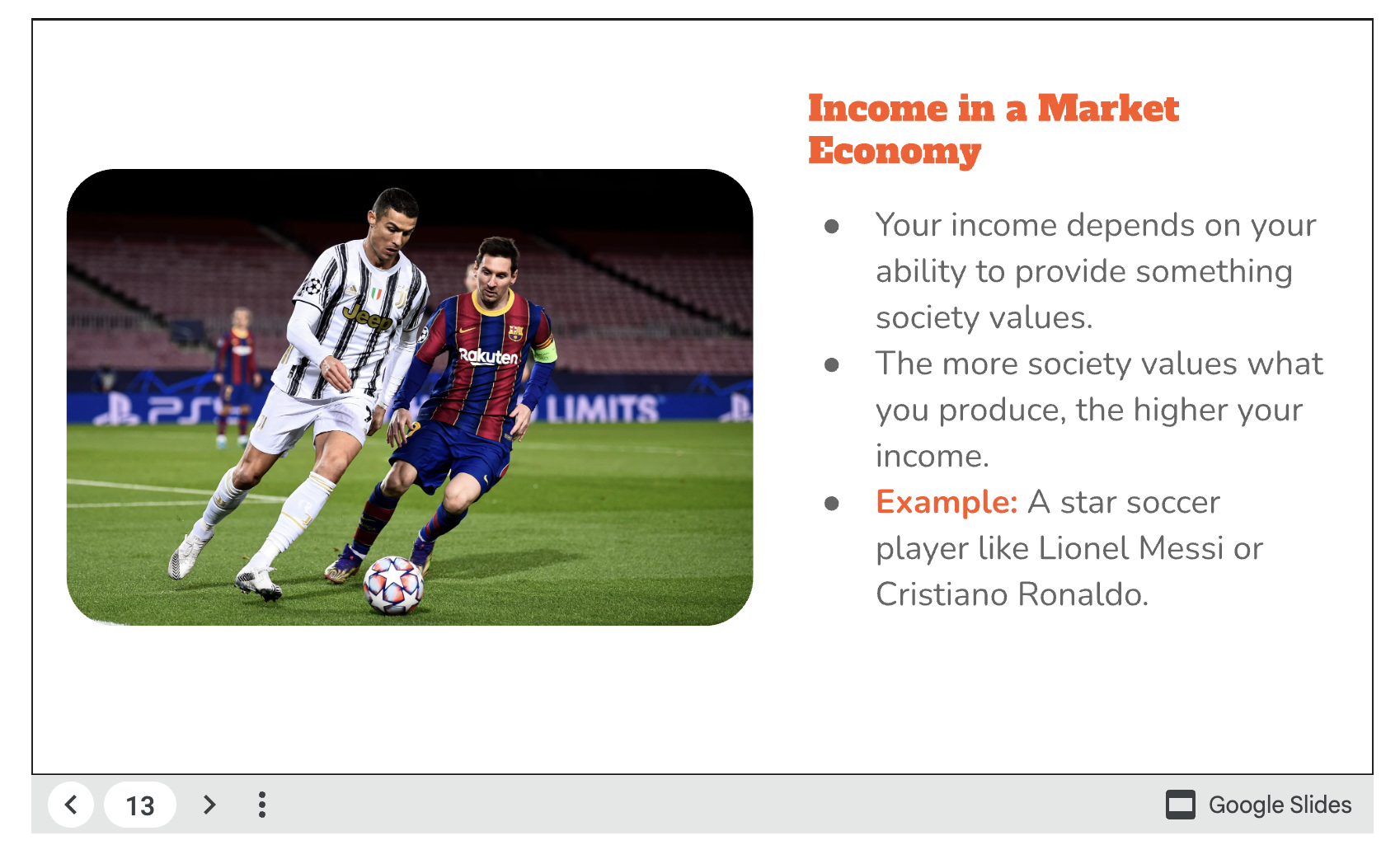}
\caption{Example slide in the deck generated for  OpenStax's \emph{How To Organize Economies} source and adapted to the learner interest in ‘soccer’.}
\label{fig:slides}
\end{figure}

\subsection{Content Transformations \label{sec:content_transforms}}
The rewriting phase above results in text that is adapted to the learner. We use this text as the basis for multiple content transformations, each providing a different view of the material. Since all are based on the personalized text in \secref{sec:text_rewrite}, they are also similarly personalized.
%
Below we provide a description of the different transformations. We describe the Slides, Audio and Mind maps transformations below, and the ``Immersive Text'' transformation in \secref{sec:immersive}.
\yh{isn't it better to talk here about the method and the rubric? Our approach when developing these capabilities is to follow pedagogical principles and learning needs. We developed a unified rubric based on the Google wide pedagogical rubric and evaluated all capabilities?}


\myHeader{Slides and Narration}
Learners often benefit from a class-like slide sequence that covers the core material in brief, while also suggesting interest-capturing questions that precede the material, and activities aimed at engagement. This provides an alternative presentation format that could be more effective for some learners. The fact that the slides are based on personalized text makes them further effective. \figref{fig:slides} shows a sample slide with an adaptation of an example of market economy to the soccer domain. 
The \xPlain{} experience also provides an additional optional generated narration for the slides. The narration is meant to resemble a recorded lesson, and the narrated text is not restricted to the text in the slides, but is rather designed to be natural and complementary to the slides.


\myHeader{Audio-Graphic lesson} 
This transformation aims at a comprehensive and detailed coverage of the material, delivered in a audio-graphic form that simulates a conversation between a teacher and a student about the material. To allow for a realistic experience, the teacher and student turns are generated iteratively using independent Gemini ``personas''. This allows for a realistic experience where the (virtual) student does not see the material before it is presented and may, for example, respond to questions with answers that are not part of the original material and uncover common learner misconceptions. In addition to the audio conversation, the lesson contains a graphical representation of the key concepts and the relationships between them, which is dynamically presented to the learner. This combination of audio and visual components is motivated by dual coding theory, which suggests that multiple representations of concepts serve to strengthen the corresponding mental conceptual structures.

\begin{figure}[t]
\centering
\includegraphics[width=1.0\columnwidth]{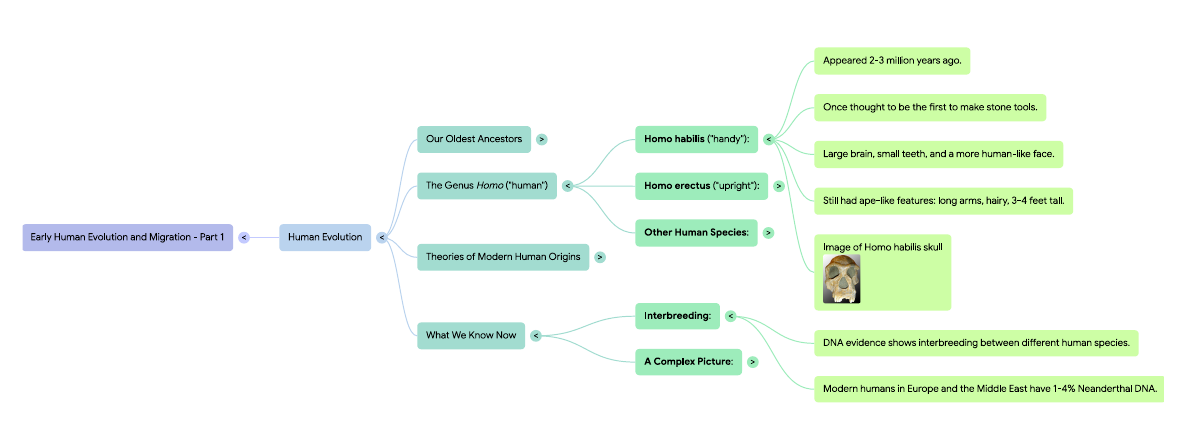}
\caption{An example mind map created for OpenStax's \emph{Early Human Evolution and Migration} source material. Different nodes can be expanded to gain a more granular view of the material, with leaf nodes annotated with text or relevant visuals.} 
\label{fig:mindmap}
\end{figure}

\myHeader{Mind Maps}
This common graphical representation organizes information hierarchically, and allows for a broad view of the material at different levels of granularity. It is often useful as a mechanism for organizing the material following a detailed learning session, or as an organizational reminder of the entire source material. We annotate the map nodes with illustrative texts and images derived from the source, and allow the user to expand and collapse nodes, in effect zooming in and out of the conceptual hierarchy. See \figref{fig:mindmap} for an example.

\subsubsection{Immersive Text \label{sec:immersive}}
The above transformations, as effective as they may be, do not stand alone. They are intended to supplement a coherent and comprehensive text articulating the content of the source in full detail. Such a text can be enhanced by interleaving it with personalized elements and multiple modalities.
We refer to this as an ``Immersive Text'' (see Figure \ref{fig:main}).
 After every section of the text we \emph{optionally} include several added components that are meant to enhance the learning experience, as detailed below. We also include assessment components, as described in \secref{sec:assessments}. 
 
\myHeader{Timeline}
Source material often contains sequences, such as a series of events in history or the stages of an experiment or algorithmic approach. ``Timelines'' can convey these sequences visually, reducing cognitive load and making it easier for the learner to follow the details. To generate these, the source material is first scanned to identify candidate sequences, followed by the generation of the timeline and appropriate placement within the material. Such transformations also offer a chance for interactive practice that is closely tied to the material: the learner is simply asked to drag-and-drop boxes into their appropriate place in the sequence.

%

\myHeader{Memory Aid}
Learning new material often involves memorizing facts, a task which can be challenging for learners. 
There are many strategies that can help learners address this need. We focus on the common strategy of mnemonics, a memorization approach where each item to remember is associated with a word that begins with the same first letter, and the sequence of words forms a sentence. With on-the-fly generation we are no longer restricted to commonly used mnemonics whose coverage is scarce. Instead, given the input material, Gemini is used to first identify elements in the material that are hard to memorize. Then a mnemonic is generated with two requirements in addition to the constraint of being a valid mnemonic: form a coherent and easy to remember sentence, and form a sentence that has close semantic association with the material to be remembered. 

\begin{figure}[t!]
\centerline{
\vspace{-0.1in}
\begin{tabular}{cc}
\begin{minipage}{0.45\columnwidth}
\vspace{-2in} {\footnotesize
A market is a system connecting buyers and sellers. The international soccer transfer market is a prime example, where clubs act as buyers and sellers of players. In this market, decisions are decentralized, and a player's income (value) is determined by how much society (clubs) values their skills.
}
\end{minipage} &
\includegraphics[width=0.45\columnwidth]{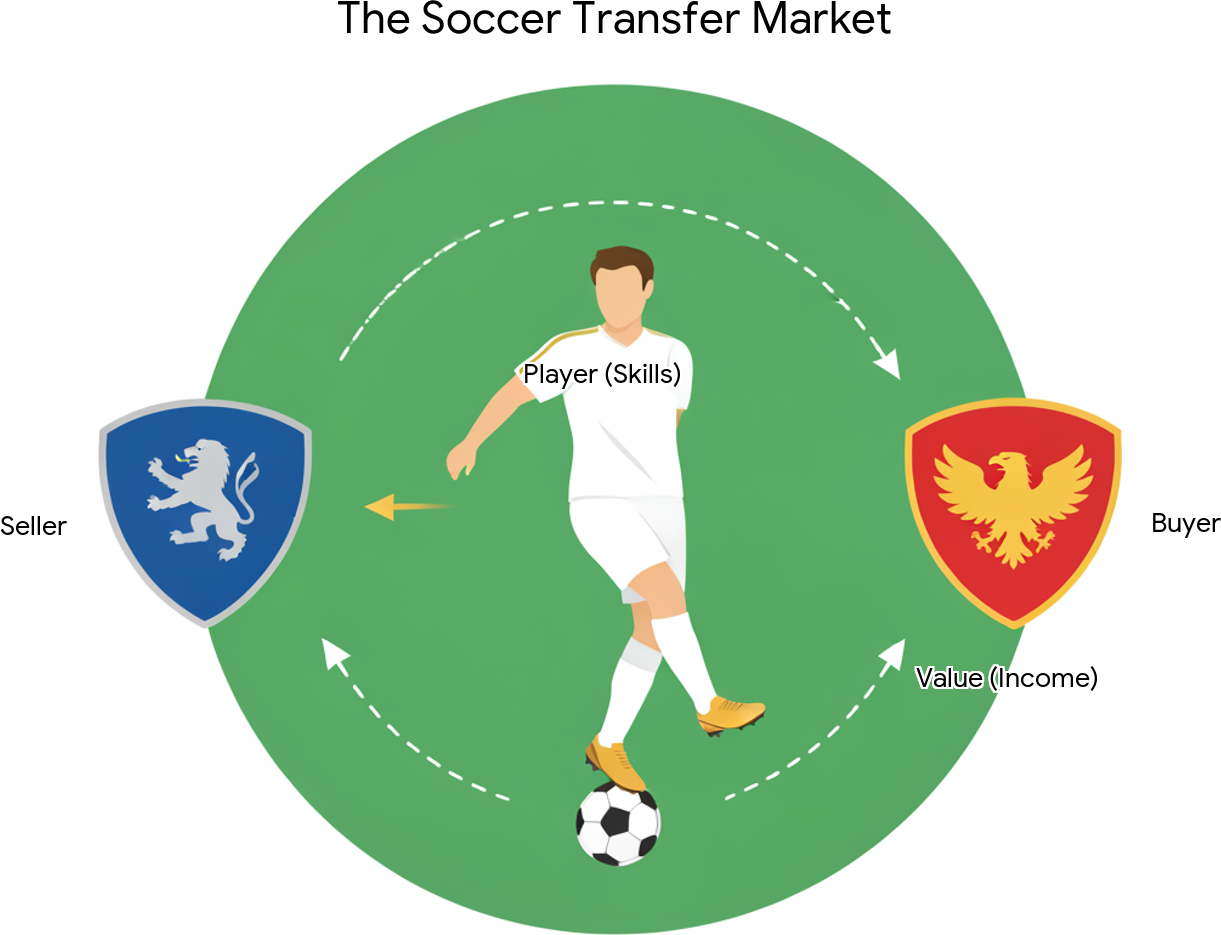}
\end{tabular}
}
\caption{An example of personalized visual generation that captures a key concept in an engaging manner that is based on the interest of the learner in soccer for OpenStax's \emph{How to Organize Economies} material. The personalized text next to the image shows that the original text has been adapted to the topic of sports. The image is a visual illustration of the text, and thus is also personalized.}
\label{fig:enimate}
\end{figure}

\myHeader{Visual Illustrations}
Visual learning is broadly recognized as a powerful medium, and many textbooks include explanatory diagrams and drawings. It is natural to use AI image generation tools to produce such visuals. However, our initial exploration found that even the most advanced AI image generation models struggle to produce these types of images. This can be explained by the fact that such models are trained to produce realistic images that are high on detail, and sometimes return inaccurate images when asked for `simple' or `educational' ones.\footnote{Note that `simple' style is not the same as simple semantics which are needed in the educational case.} To overcome this, we fine-tuned a model specifically for this task. This model is applied to parts of the material that Gemini identifies as worthy of illustration. An example, is shown in \figref{fig:enimate}.

\section{Practice and Assessment \label{sec:assessments}}
Formative assessment is arguably one of the primary drivers of learning \cite{black2009developing}, and augments the multiple views of material discussed previously. Indeed, effective while-you-learn assessment and on-the-fly feedback can help reinforce concepts and increase knowledge and skill retention \citep[e.g., see][]{earl2013assessment}. We therefore augment \xPlain{} with two assessment components, described below. Both components appear within the Immersive Text view.



\begin{figure}[t]
    \centering
    \begin{tabular}{cc}
    \begin{minipage}{0.42\columnwidth}
        \vspace{-2in}
        \caption{An example of an embedded question for OpenStax's topic of \emph{How to Organize Economies}.}
    \label{fig:mnemonics_and_question}
    \end{minipage} &
    \includegraphics[width=0.52\columnwidth]{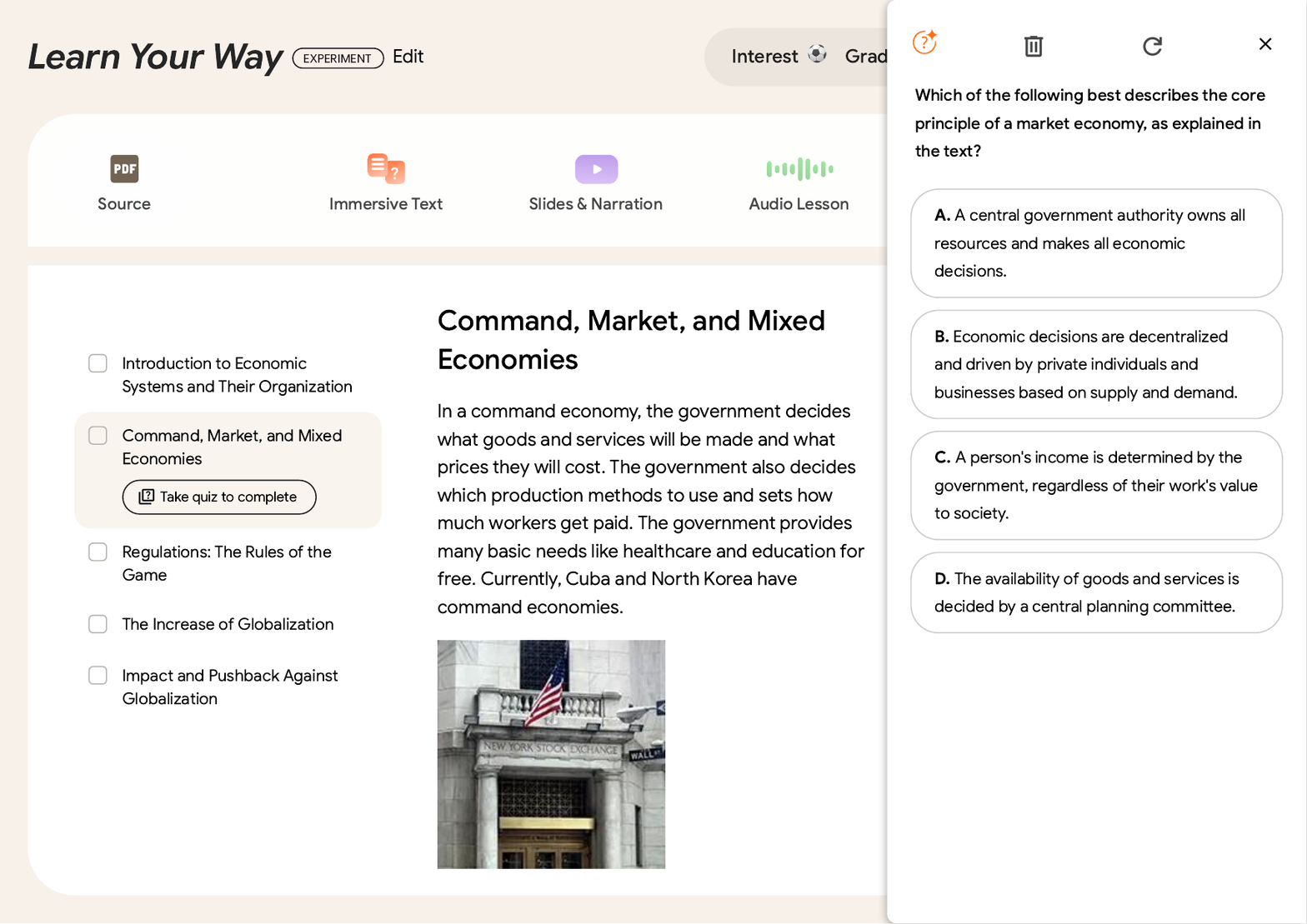} 
    \end{tabular}
\end{figure}

\myHeader{Embedded Questions}
Embedded questions are dynamically generated questions that are grounded and associated with specific segments of the source material. These questions serve to convert the reading  experience from passive to active and to keep the learner engaged by providing immediate feedback. They also reinforce the concept being learned. In \xPlain{} they are presented as multiple choice questions that appear when the learner clicks a question mark in the Immersive Text view. See \figref{fig:mnemonics_and_question}.

\myHeader{Quizzes}
Section-level quizzes aim at deeper understanding once a section has been read and assimilated. The quizzes are dynamically generated and grounded to all of the material in the section. They consist of 5-10 multiple choice questions of various difficulties and types. At the end of the quiz, an overall assessment is provided that includes both a numerical score as well as targeted feedback that highlights strengths (or Glows) and areas for improvement (or Grows). 

\section{Pedagogical Evaluations}
In order to assess the quality of the different augmentation and transformation components used in \xPlain{}, we asked multiple pedagogical experts to evaluate the quality of each according to a pedagogical rubric over several criteria. 

For the evaluation we used ten source-of-truth PDFs, ranging in topics from sociology to physics from \href{https://openstax.org/}{OpenStax} (see full list in \appref{app:materials}). For grade level personalization, we considered three grade levels (7th grade, 10th grade and undergraduate level) and three personal interests (basketball, music and food). Each PDF was assigned three random combinations of grade level and personal interests (out of the nine possible combinations). 
%
Each of the configurations above was then provided as input to \xPlain{}, which generated the transformations and assessments described in \secref{sec:content_transforms} and \secref{sec:assessments}.
The experts were asked to evaluate the overall merit of the \xPlain{} experience, as well as the individual components of the system:
\begin{itemize}
    \item Content transformations: Slides, Narrated Slides, Audio Lesson, Mind map. 
    \item Immersive text components: Timeline, Memory Aid, Visual illustrations.
    \item Personalization to Interests: as noted in \secref{sec:text_rewrite}, parts of the original text were replaced with a version personalized to the interests of the user. These parts were also highlighted, and thus we could ask the pedagogical raters to evaluate them.
    \item Assessment: as noted in \secref{sec:assessments}, the immersive text contains two types of assessments: embedded questions and quizzes. We evaluate these separately.
\end{itemize}

\begin{figure}[t]
    \centering
    \begin{tabular}{cc}
    \includegraphics[width=\linewidth]{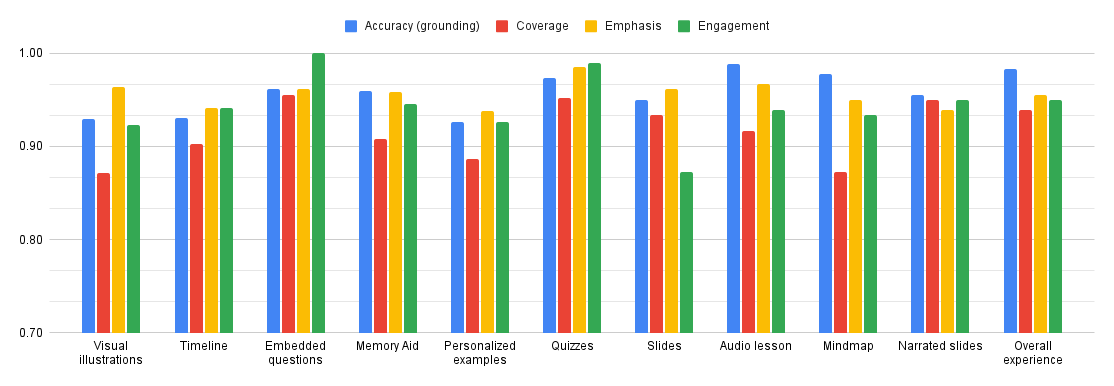} \\
    \includegraphics[width=\linewidth]{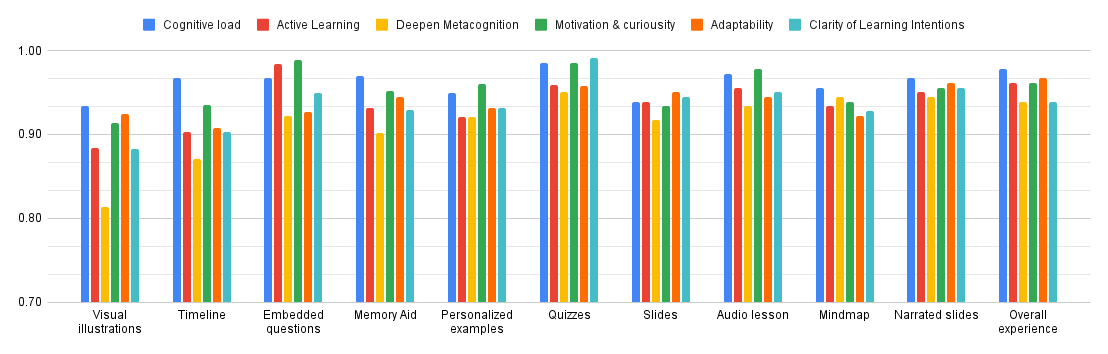} \\
    \end{tabular}
    \caption{Rating of the various components that make up \xPlain{}, as rated by pedagogy experts. Top figure shows high-level metrics, and the bottom figure shows additional metrics that arise from core learning science principles.
    Rating is based on the rubrics detailed in \appref{app:rubrics}. Shown is the average across the three raters across all source materials.}
    \label{tab:pedagogy}
\end{figure}

We asked the pedagogical experts to evaluate the quality of each component with respect to several basic criteria such as coverage, as well as criteria that capture key learning science principles.
See \appref{app:rubrics} for the evaluation rubrics for all criteria. For example, for the Accuracy criterion, they were asked if the component is ``faithful to the source and accurate''. The criteria were Coverage, Emphasis, Cognitive load, Active learning, Deepen meta-cognition, Motivation \& curiosity, Adaptability, and Clarity of learning intentions. For each of these, following the guidance of pedagogical rubrics (see \appref{app:rubrics}), raters provided an agree ($1.0$), neutral/partial ($0.5$), or disagree ($0.0$) rating. Finally, each component was evaluated by three different raters. 

The results are summarized in \figref{tab:pedagogy}, which shows that all components have relatively high pedagogy values and the overall experience is rated over $0.90$ across all axes. The component with the lowest scores is that of Visual Illustration. This is to be expected given the difficult of generating high quality pedagogical images. 


A closer inspection of the results presented in \figref{tab:pedagogy} reveals additional more nuanced insights. For example, the slides format received the lowest `engagement' score of all capabilities. On the other hand, these same slides but with generative narration received a significantly higher score. This is in line with the fact that slides are often presented alongside narration, and thus this combination is more engaging for learners. These types of insights are invaluable for improving future versions of all \xPlain{} components.

\section{Efficacy Study Design and Results
\label{sec:rct_main}}

\begin{figure}[t]
    \centering
    \vspace{-0.1in}
    \includegraphics[trim={0.05cm 0 0 0},clip,width=1.0\linewidth]{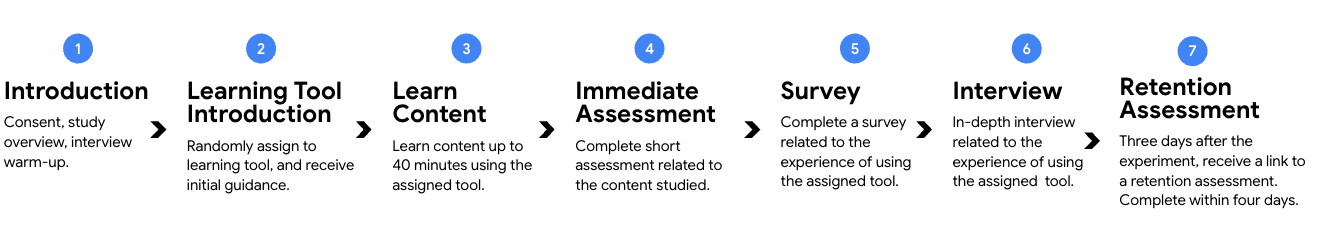}

    \caption{Overview of the experimental setup of the personalized learning efficacy study. 
    }
    \label{fig:rct_protocol}
\end{figure}

Above we reported on a pedagogical evaluation of the different generated components for multiple topics and personalizations. However, the impact is much greater when the various capabilities come together in a learning experience, and it is therefore important to measure their holistic pedagogical value in terms of learning efficacy. To evaluate \xPlain{} from this perspective, we ran an experiment where students had to learn an unfamiliar textbook chapter.

\myHeader{Student Recruitment} Sixty students, aged 15-18 years old, were recruited from the Chicago area across urban, suburban, and rural schools.
To ensure that participating students had similar reading comprehension skills, we gave them a reading comprehension and assessment task as part of the recruitment criteria. The assessment included a short passage followed by a series of questions about the passage. As such, the format of this recruitment task was similar to the experiment, but with different learning material (it was a text on ocean waves). The average score on the assessment was 6.4, with a standard deviation of 2.3. We included students who scored 1 standard deviation above or below the mean, which corresponded to receiving a score of 4-9 out of 10 on the assessment.

\myHeader{Learning Material and Assessment}
In our study, all students were presented with the same textbook chapter (Brain Development for Adolescents from \href{https://socialsci.libretexts.org/Bookshelves/Human_Development/Lifespan_Development_(Lumen)/07%3A_Adolescence/7.04%3A_Brain_Development_During_Adolescence}{LibreTexts}). 
In choosing the material, a panel of pedagogy experts was asked to find a text that conformed to several pedagogical criteria, including length, familiarity, and interest. To account for pre-study familiarity with the topic in our analysis, students were asked to record their familiarity with the topic before the study on a scale of $1-5$ with $5$ corresponding to very familiar.

After exploring the learning material, all students received the same assessments, which were written by a pedagogical expert.
%
Two assessments were written by a pedagogical expert to evaluate understanding of the learning material. The first ``Immediate Assessment'' was a 15 minute assessment designed to be taken immediately after the learning session. This included short answer questions, single-answer, multiple choice questions, matching questions, multi-answer, and multiple choice questions. The questions were written by the expert to match Bloom's taxonomy level expected at that age group. Three days post-session, participants were emailed a follow-up ``Retention Assessment'' that measured long-term recall of the content they had learned during the experiment. It was designed to take $5-10$ minutes, and included a short answer question, a single-answer multiple choice question, and a matching question. Participants were given four days to complete it.  58 out of 60 participants completed the retention assessment. 

Immediately after the first assessment, students were also asked to rate its difficulty. Students using \xPlain{} rated the assessment's difficulty as $3.6$ on average (on a scale of $1-5$), slightly higher than those in the control group who gave an average rating of $3.37$, but the difference is not statistically significant between the groups using the Mann-Whitney U test. These values suggest that, from the learners' perspective, the assessment was neither too difficult nor too easy.

\myHeader{Experimental Setup} Participants were brought into the lab and introduced to the study via informed consent, followed by an overview of the research sessions. Each participant was randomly assigned to one of two learning conditions: \xPlain{} and a Digital Reader (Adobe Acrobat Reader v25.001.20531). For students assigned to \xPlain{}, the average reported pre-study familiarity with the topic was $2.6 \pm 1.1$.  Students who were assigned to the digital reader were slightly more familiar with the topic with an average of $2.9 \pm 1.2$, but the difference was not found to be statistically significant using the Mann-Whitney U test.

Each participant had five minutes to review a set of three slides that introduced them to the features available in the tool they were assigned. Participants then used the assigned tool to study the material. Learning time was set to a minimum of 20 minutes and a maximum of 40 minutes. After this time, each participant had 15 minutes to complete the Immediate Assessment via a Qualtrics link. Participants did not receive a visible score after completing the assessment. Each participant then had 10 minutes to answer a series of quantitative questions via Qualtrics to share their impressions of the learning tool. The moderator followed-up with a 20 minute 1:1 qualitative interview with each participant, diving deeper into their experience with the tool. 
Three days after the study, participants received the Retention Assessment, as explained above. The setup is summarized in \figref{fig:rct_protocol}.

\begin{figure}[t] \centering
\begin{tabular}{cc}
\includegraphics[width=0.48\columnwidth]{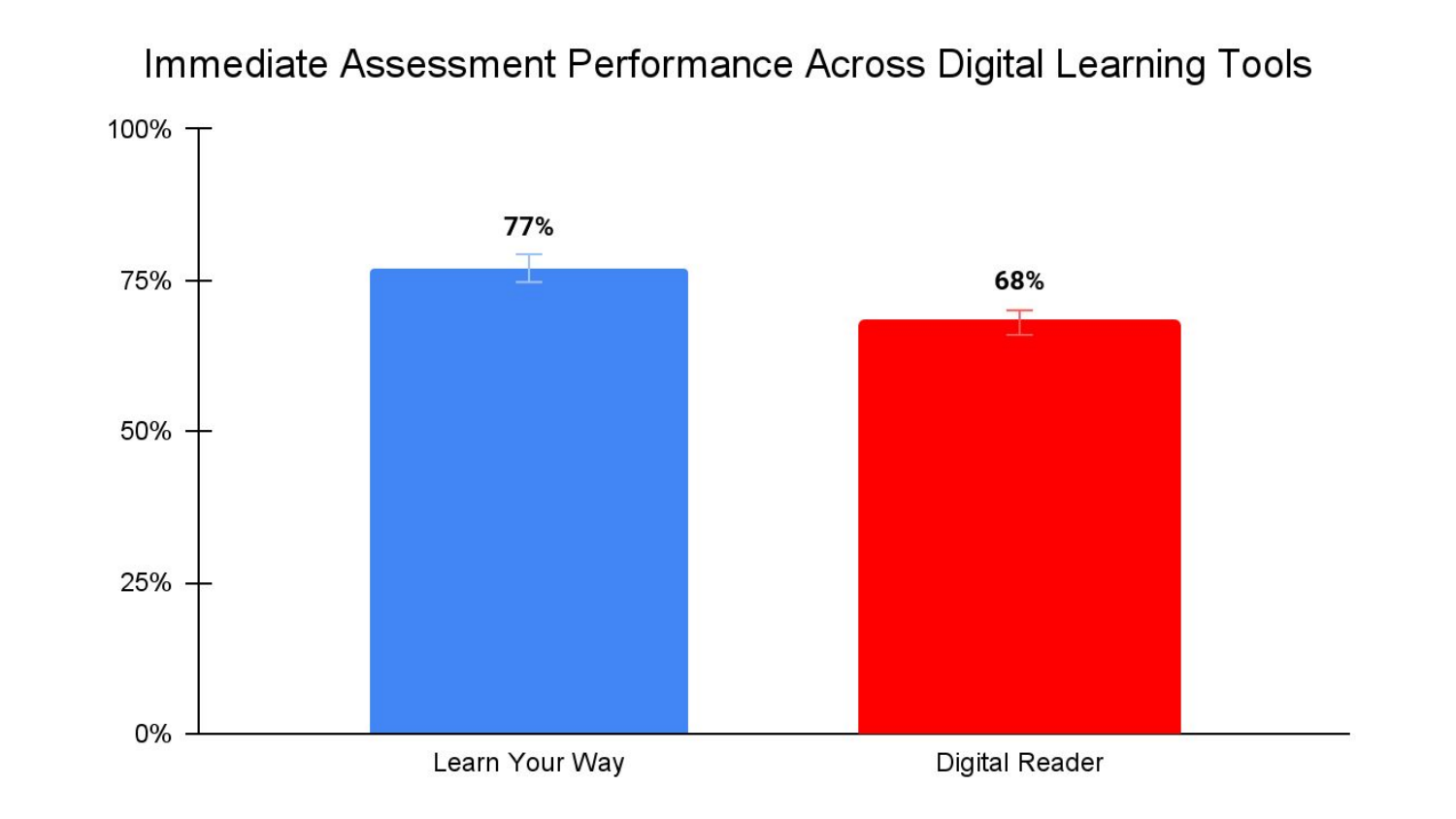} 
\includegraphics[width=0.48\columnwidth]{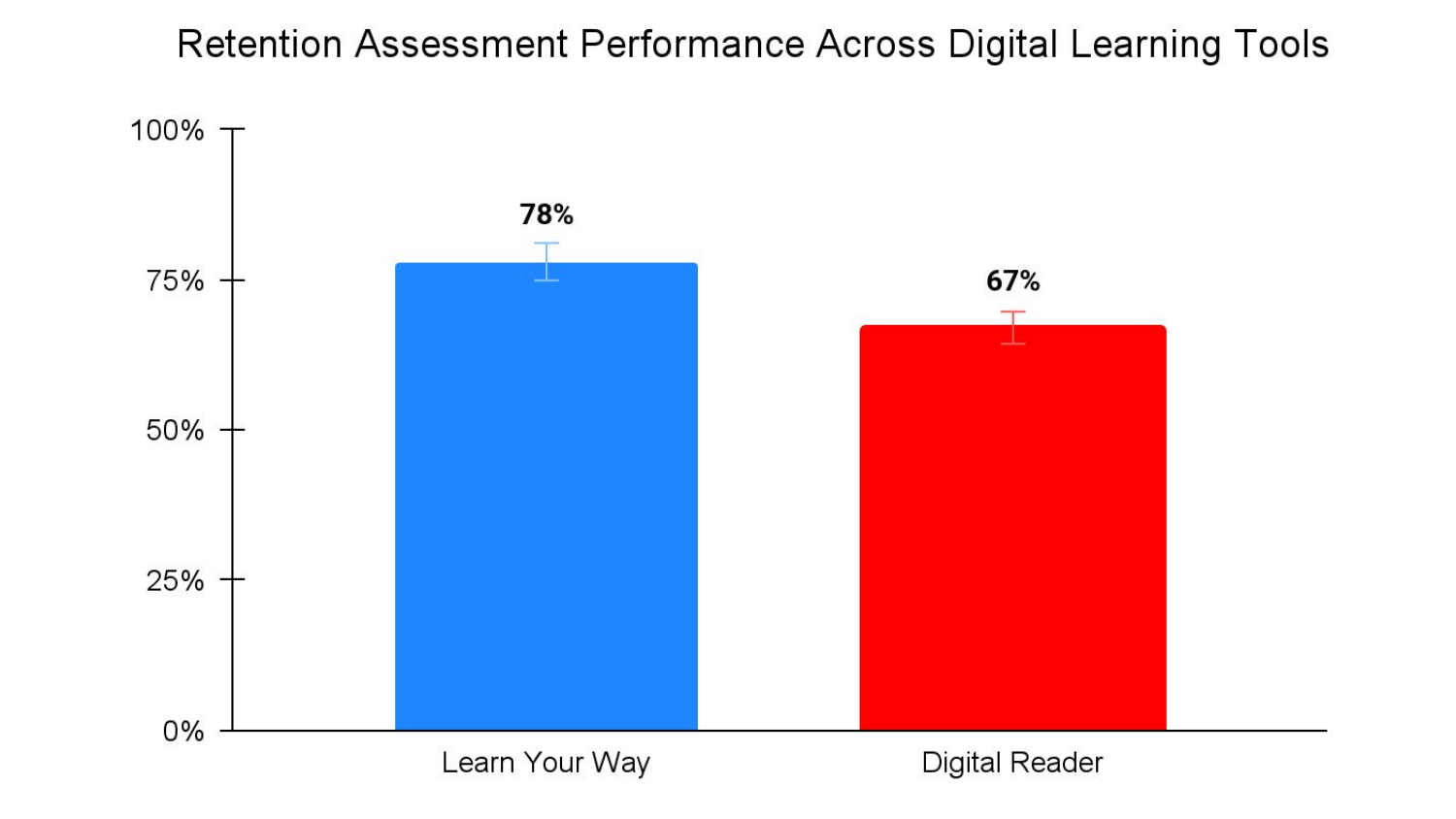} 
\end{tabular}
\caption{(left) Average scores of learners who used \xPlain{} and a  Digital Reader for the assessment that immediately followed the learning experience. (right) Average scores for the retention assessment given three days after the learning experience. In both cases, the differences between \xPlain{} and the tool were statistically significant.}
\label{fig:performance}
\end{figure}

\myHeader{Results} 
\figref{fig:performance} (left) shows the results from the Immediate Assessment, and \figref{fig:performance} (right) shows results from the Retention Assessment.
 The students who used \xPlain{} received higher scores than those who used the Digital Reader, in both the immediate (\textit{p} = 0.03) and retention (\textit{p} = 0.03) assessments. The Shapiro-Wilk test was used to test normality of the scores and, since this was not the case, the nonparametric Mann-Whitney U test was used to compute statistical significance. 
 

\begin{figure}[t]
    \centering
    \includegraphics[trim={35 80 35 80},clip,width=1.\linewidth]{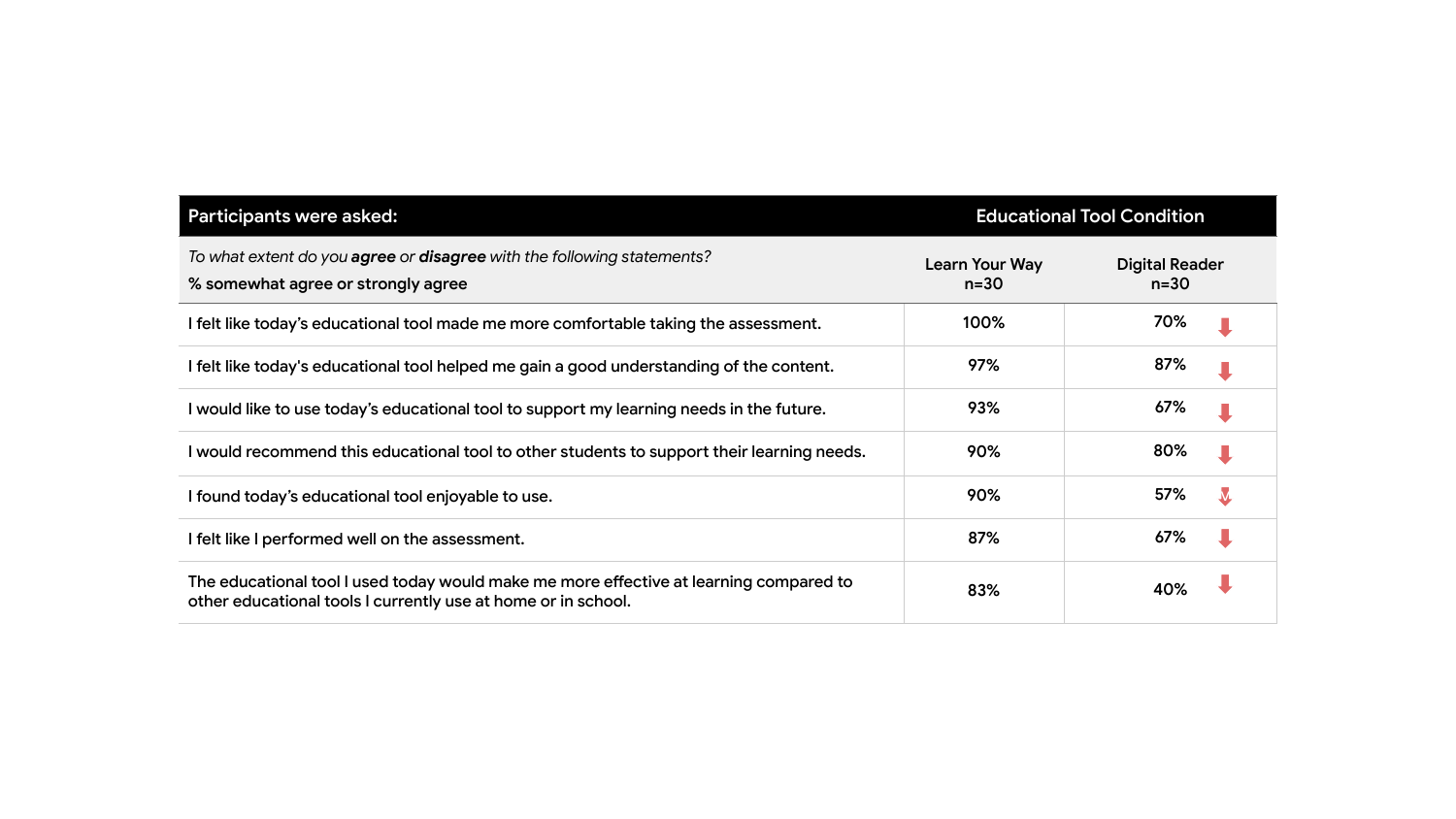}
    \caption{Learner responses to a survey given after the assessment. For each statement, shown is the percentage of students who agreed or strongly agreed with it for each cohort. Red down arrow indicates that \xPlain{} had significantly higher ratings than the Digital Reader (\textit{p} < .05).} \vspace{0.1in}
\label{fig:responses}
\end{figure}

\figref{fig:responses} reports results on the learning experience survey. It can be seen that that \xPlain{} provides significant advantages compared to the Digital Reader. Specifically, across all measures that assessed learning experiences, \xPlain{} was consistently evaluated more positively compared to the Digital Reader using the same statistical tests as above.


\section{Discussion and Future Work}
We presented a two stage Gen-AI approach for transforming and augmenting source learning material. Our approach was implemented using Gemini in \xPlain{}, a novel, experimental learning experience. In addition to pedagogical expert evaluation of the quality and merit of the different components of \xPlain{}, a randomized controlled trial substantiated its potential efficacy for real student learning given an unfamiliar textbook chapter.

The scope of our efficacy study has  limitations. Since \xPlain{} contains multiple components, including formative quizzes, a natural question is which of these contributes most to learning efficacy. Since the current study did not hone in on this, there might be some transformations that have impact while others do not. It would also be beneficial to broaden the evaluation to multiple textbook chapters and topics. Addressing these questions within a more detailed controlled experiment is possible, but would be costly. On the other hand, analysis of sessions of learners who use \xPlain{} can provide extensive information, and we plan to pursue this in the future. For example, we recorded which \xPlain{} components were used by the different subjects. One interesting observation  was that the majority of participants in the study used at least one content transformation in addition to the immersive text, and all made use of quizzes during the learning session.  



This work is the tip of the iceberg of the potential impact of generative AI on learning at large and learning personalization in particular. The \xPlain{} experience could be extended in many ways, and the element of personalization could be explored and pushed much further. Additional learner attributes could be researched, through both explicit and implicit signals. The system could be made more adaptive, by dynamically adjusting the learning material to the performance of the learner on assessment components (e.g., to focus on learning gaps), or to the individual needs and difficulties of the learners. More interactive elements could be added to increase student interactions, both to obtain more signals from the learner and to create a more effective learning experience.
\xPlain{} could be embedded in learning platforms, in ways that will provide teachers with control and insights into the learning process. The above feedback processes could then be designed to optimize learning efficacy.

Above all, \xPlain{} demonstrates how the imaginative use of generative AI, grounded in solid learning science principles and crafted and evaluated with pedagogical experts, opens up exciting opportunities to enhance learning.



\bibliography{plug}

\section*{Contributions and Acknowledgments}

\myHeader{Core Contributors}
The following individuals made core contributions to the work described in this report. This list is
ordered alphabetically, and does not indicate ranking of contributions:

Alicia Martín,  Amir Globerson,  Amy Wang, Anirudh Shekhawat, Anisha Choudhury,  Anna Iurchenko,  Avinatan Hassidim,   Ayça Çakmakli, Ayelet Shasha Evron,  Charlie Yang,  Courtney Heldreth,  Diana Akrong,  Gal Elidan,  Hairong Mu,  Ian Li,  Ido Cohen,  Katherine Chou, Komal Singh,  Lev Borovoi,  Lidan Hackmon,  Lior Belinsky,  Michael Fink,  Niv Efron,  Preeti Singh, Rena Levitt,  Shashank Agarwal,  Shay Sharon,  Tracey Lee-Joe,  Xiaohong Hao,  Yael Gold-Zamir,  Yael Haramaty,  Yishay Mor,  Yoav Bar Sinai,  Yossi Matias

\myHeader{Acknowledgments}
We completed this work as part of the LearnLM effort—a cross-Google project, with members from Google DeepMind, Google Research, Google LearnX, and more. This tech report represents only a
small part of the wider effort, and only lists team members who made direct contributions to this
report.
Special thanks to Ben Gomes, Irina Jurenka, James Manyika, Julia
Wilkowski and Muktha Ananda for invaluable feedback.

\appendix
\section{Source Materials}
\label{app:materials}
The PDFs used as the souce-of-truth for the pedagogical evaluations were all thanks to \href{https://openstax.org/}{OpenStax}. The 10 varied PDFs used are listed in the table below:

\begin{table}[ht]
\begin{tabular}{|l|l|}
\hline
Category & Title \\
\hline
World History &	\href{https://openstax.org/books/world-history-volume-1/pages/2-1-early-human-evolution-and-migration}{Early Human Evolution and Migration} \\
World History	& \href{https://openstax.org/books/world-history-volume-1/pages/7-3-the-roman-economy-trade-taxes-and-conquest}{The Ancient Roman Economy} \\
World History & 	\href{https://openstax.org/books/world-history-volume-2/pages/14-1-the-cold-war-begins}{The Cold War Begins} \\
Biology	& \href{https://openstax.org/books/biology-2e/pages/26-1-evolution-of-seed-plants}{Evolution of Seed Plants} \\
Biology	& \href{https://openstax.org/books/biology-2e/pages/42-4-disruptions-in-the-immune-system}{Disruptions in the Immune System} \\
Physics	& \href{https://openstax.org/books/physics/pages/4-4-newtons-third-law-of-motion}{Newton's Third Law of Motion} \\
Economy	& \href{https://openstax.org/books/principles-economics-3e/pages/1-4-how-to-organize-economies-an-overview-of-economic-systems}{How To Organize Economies: An Overview of Economic Systems} \\
Astronomy &	\href{https://openstax.org/books/astronomy-2e/pages/3-5-motions-of-satellites-and-spacecraft}{Motions of Satellites and Spacecraft} \\
Sociology &	\href{https://openstax.org/books/introduction-sociology-3e/pages/5-1-theories-of-self-development}{Theories of Self-Development} \\
Psychology &	\href{https://openstax.org/books/psychology-2e/pages/4-2-sleep-and-why-we-sleep}{Sleep and Why We Sleep} \\
\hline
\end{tabular}
\end{table}
\clearpage
\section{Rubrics for Pedagogical Evaluations}
\label{app:rubrics}

\begin{table}[h!]
{\tiny
\begin{tabular}{|p{1.3cm}|p{1.8cm}|p{4cm}|p{3.5cm}|p{3.5cm}|}
\hline
 & & AGREE (1) & NEUTRAL / PARTIAL (0.5) & DISAGREE (0) \\ \hline
Accuracy & Faithful to the source and accurate & The generated content is consistent with the source resource. Does not misrepresent or misconnect concepts. Captures and clearly conveys the source's main arguments, supporting key claims. The organization of the content (when applicable) is accurate and factual.  & Content does not add or alter main concepts, but might have minor inaccuracies, or miss concepts, relationships, or arguments.  & Content contains claims or arguments that counter the source, major inaccuracies, or misses major concepts. \\ \hline
Coverage & Completeness of representation of source messages & The generated content covers the source content, faithfully preserving the links and structure of the original. The source hierarchy is preserved. All key / top level concepts and LOs are covered, but immaterial details may be omitted to support learnability.   For global constructs - global coverage.   For local constructs - local coverage, i.e. projected to the scope covered by the construct. & Content mostly covers the structure and subject matter of the source, but is missing some key concepts, relationships, or messages. Source hierarchy is not preserved, some immaterial details are carried over while more important elements omitted. & Key subject matter or structure is missing from the generated content.  \\ \hline
Emphasis & Precisely prioritizes core concepts. Identifies the source hierarchy, and optimal chooses and places key messages. & The generated content focuses on the key concepts / relationships in the source, not trivial / marginal / esoteric concepts.   For global constructs: The hierarchy of elements correlates with the knowledge / skills hierarchy of the source.   For local constructs: The generated content focuses on the prominent elements in the source. & Choice of elements partially reflects source hierarchy or intentions by including most of the core concepts, but might miss an important concept or emphasis some trivial material along with the core concepts.  & Choice of elements appears arbitrary, and does not align with source hierarchy or intents. Generated content highlights immaterial concepts or relationships instead of key messages. \\ \hline
Engagement & Positive, pleasant and purposeful user experience & The generated content evokes positive emotional response. The tone, aesthetics, narrative, playfulness, and elements of personalization create a fun and captivating learning experience.  & Content does not consistently engage or evoke a positive response from the user. Limited use of playfulness or personalization. Utility and relevance to user is present, but might not be clear to the user. & Content is dry, boring, does not attempt to evoke positive emotions. Monotonous single-track presentation. No perceived value to the user. \\ \hline
Cognitive load & Effective use of cognitive effort & Content is well organized, clear and concise. Accessible language. Key points emphasized. Uses examples, analogies, narrative, and multiple representations effectively to enhance understanding.  & Content is fairly well structured but some sections could be improved. Phrasing is occasional difficult. Some unhelpful redundancy. Limited use of examples, analogies, narratives, or multiple representations to enhance learning. & Content is poorly structured, and may included large overwhelming blocks of text. No clear hierarchy. The language level is inconsistent, incorporating difficult terms or oversimplifying for the users level. No use of aids such as examples, analogies, narratives, or multiple representations. \\ \hline
Active Learning & Promotes active learning that goes beyond recall and comprehension. & Content raises questions and encourages user to engage with subject matter and learning objectives in multiple levels: recall, comprehension, analysis, application, evaluation, and creation. & Content mostly focuses on information delivery, and misses opportunities to engage user with learning objectives at multiple levels. & Content is purely informational and does not encourage higher order thinking. \\ \hline
Deepen Metacognition & Promotes active self-reflection, monitoring, regulation, and improvement of thinking and learning processes. & Provides relevant and actionable feedback. Prompts active reflection and inspection of learning processes. Supports the user to identify, execute and monitor their learning plan.  & Provides somewhat useful feedback, but may not be relevant or actionable. Prompts shallow reflection without consideration of learning processes, or limited consideration of learning plans. & Flat informative content, no useful feedback, no prompts for reflection or managing and regulating learning. \\ \hline
Motivation \& curiosity & Encourages users to persist and deeply engage with learning processes. Incentivizes cognitive effort. & Uses a consistently supportive tone, encouraging user to persist. Maintains an optimal level of challenge, between too easy (boring) and too hard (frustrating). Highlights the relevance of the subject matter to the user, their concerns and interests. Empowers user autonomy by clearly marking available choices of learning depth, pace, order and focus. & Tone is occasional unsupportive, or encouragement is sporadic. Challenges are not consistently aligned to an optimal challenge level  for the user. Limited use of examples and real-life connections to highlight subject matter relevance to user's concerns and interests. User choices in the learning process are not evident. & Tone is not supportive or encouraging. Challenges are missing, or at inappropriate level for the user. No evident links from the subject matter to user concerns and interests. No choice in learning path. \\ \hline
Adaptability \& Personalization & Learning experience is adjusted to fit the user interests, preferences and characteristics. & Content is fully adjusted to user's interests, age group, grade level, language proficiency, level of focus and attention, motivation, and preferences. Challenges are appropriate for the user's capabilities and intents. & Content acknowledges user's preferences and characteristics, but is only partially adjusted. Challenges are not consistently aligned with user capability or intent. & Content ignores user preferences and characteristics, "one size fits all". \\ \hline
Clarity of Learning Intentions \& Success Criteria & Content clearly articulates what is being learned (learning intentions/goals/objectives) and how users will know if they have been successful (success criteria).  & Learning objectives are clearly stated and user-friendly. Specific rubrics or benchmarks define what successful completion or understanding looks like. Content directly aligns with stated LOs and criteria (if provided). & Learning objectives stated but not elaborated or demonstrated. Success criteria are vague or confusing for the user. Content alignment with LOs partial or unintuitive (if provided). & Learning objectives not specified, unclear or inconsistent. No evident success criteria. Content misaligned with LOs (if provided). \\ \hline
\end{tabular}
} 
\caption{Pedagogical rubrics used by experts to rate the various components of \xPlain{}. Raters were also given the option of marking \emph{N/A or can't assess} for each dimension.}
\label{tab:rubrics}
\end{table}

\end{document}